\documentclass[12pt,epsf,epsfig,psfig]{article}
\usepackage{graphicx}
\usepackage{epsfig}
\usepackage{cite}
\def\q{q{\bar q}}

\def\e{\rm e}

\def\n{n_{\rm cl}}

\def\n{\bar n}
\def\w{\rm vdW}

\def\be{\begin{equation}}
\def\ee{\end{equation}}
\def\be{\begin{equation}}
\def\ee{\end{equation}}

\def\lsim{\raise0.3ex\hbox{$<$\kern-0.75em\raise-1.1ex\hbox{$\sim$}}}
\def\gsim{\raise0.3ex\hbox{$>$\kern-0.75em\raise-1.1ex\hbox{$\sim$}}}

\def\NP{{ Nucl.\ Phys.\ }}
\def\PL{{ Phys.\ Lett.\ }}
\def\PR{{ Phys.\ Rev.\ }}

\def\PRL{{ Phys.\ Rev.\ Lett.\ }}

\def\ZP{{ Z.\ Phys.\ }}

\begin{document}

~\hfill BI-TP 2008/17

\vskip1cm

\centerline{\Large \bf The Phase Diagram of Hadronic Matter}

\vskip 1cm

\centerline{\large \bf P.\ Castorina$^a$, K.\ Redlich$^{b,c}$ and H.\ Satz$^d$}

\medskip

\centerline{$^a$ Dipartimento di Fisica, Universit{\`a} di Catania,
and INFN Sezione di Catania, Italy}

\centerline{$^b$ Institute of Theoretical Physics, University of Wroc{\l}aw,
Poland}

\centerline{$^c$ Technische Universit{\"a}t Darmstadt, 
Germany}
                    
\centerline{$^d$ Fakult{\"a}t f{\"u}r Physik, Universit{\"a}t Bielefeld, 
Germany}

\vskip 1cm

\centerline{\large \bf Abstract}

\medskip

We interpret the phase structure of hadronic matter in terms of 
the basic dynamical and geometrical features of hadrons. Increasing 
the density of constituents of finite spatial extension, by increasing 
the temperature $T$ or the baryochemical potential $\mu$, eventually 
``fills the box'' and eliminates the physical vacuum. We determine the 
corresponding transition as function of $T$ and $\mu$ through 
percolation theory. At low baryon density, this means a fusion of
overlapping mesonic bags to one large bag, while at high baryon
density, hard core repulsion restricts the spatial mobility of
baryons. As a consequence, there are two distinct limiting regimes for
hadronic matter. We compare our results to those from 
effective chiral model studies.

\section{Introduction}

\begin{figure}[htb]
\centerline{\psfig{file=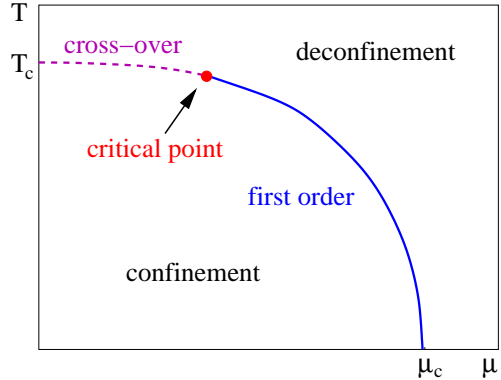,width=6.5cm}}
\caption{Phase structure of QCD matter in the $T-\mu$ plane}
\label{phase}
\end{figure}

In recent years, the general features of the phase diagram of strongly
interacting matter have become increasingly well established \cite{Karsch}. 
Lattice QCD studies at finite temperature and now also for some range 
of finite baryon density \cite{Fodor,Forcrand,Lombardo,Swansea}, combined 
with chiral symmetry restoration arguments \cite{Stepha,Halasz,SRS,SKP,
Buballa,WW,CS}, lead for physical quark masses to the phase structure 
illustrated in Fig.\ \ref{phase} as function of temperature $T$ and 
baryochemical potential $\mu$. Following a region of non-singular but 
rapid cross-over of thermodynamic observables around a quasi-critical 
temperature of 170 - 190 MeV, increasing $\mu$ leads to a critical point, 
beyond which the system shows a first order transition from confined to 
deconfined matter.

\medskip

\medskip

The aim of the present work is to deduce such behaviour from the basic
dynamical and geometrical features of hadronic matter. Our question  
thus is the following: what conceptual aspects of hadronic interactions 
lead to the observed behaviour, and in particular, what features in 
hadronic dynamics result in the observed changes of the transition structure
as function of $T$ and $\mu$?

\medskip

At low baryon density, the constituents of hadronic matter are mostly
mesons, and the dominant interaction is resonance formation; with increasing 
temperature, different resonance species of increasing mass are formed, 
leading to a gas of ever increasing degrees of freedom. They are all of 
a typical hadronic size (with a radius $R_h \simeq 1$ fm) and can 
overlap or interpenetrate each other. For $\mu \simeq 0$, the contribution 
of baryons/antibaryons and baryonic resonances is relatively small, but with 
increasing baryon density, they form an ever larger section of the species 
present in the matter, and beyond some baryon density, they become the 
dominant constituents. Finally, at vanishing temperature, the medium 
consists essentially of nucleons.

\medskip

\begin{figure}[htb]
\centerline{\psfig{file=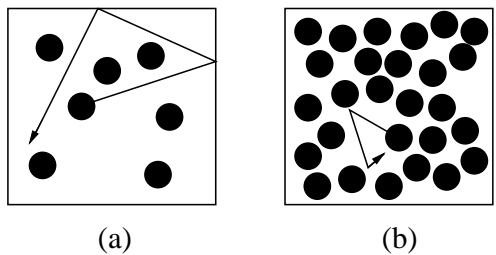,width=6.5cm}}
\caption{Hard sphere states: full mobility (a), ``jammed'' (b)}
\label{hard}
\end{figure}

At high baryon density, the dominant interaction is non-resonant. 
Nuclear forces are short-range and strongly attractive at distances of 
about 1 fm; but for distances around 0.5 fm, they become strongly repulsive. 
The former is what makes nuclei, the latter (together with Coulomb and Fermi 
repulsion) prevents them from collapsing. The repulsion between a proton 
and a neutron shows the purely baryonic ``hard-core'' effect and is connected 
neither to Coulomb repulsion nor to Pauli blocking of nucleons. As a 
consequence, the volumes of nuclei grow linearly with the sum of its 
protons and neutrons. With increasing baryon density, the mobility of 
baryons in the medium becomes strongly restricted by the presence of other 
baryons, leading to a ``jammed'' state, as shown in Fig.\ \ref{hard} 
\cite{KS}. 

\medskip

\medskip

Increasing the density of constituents by increasing temperature, baryon 
density or both, leads to clustering of hadrons of spatial extension,
and these clusters will eventually span the entire available volume. 
This onset of connectivity and the associated geometric critical behaviour
is the central topic of percolation theory, where the relevant thresholds
have been determined for permeable spheres which can overlap (mesons) as well 
as for those which have a impenetrable hard core (baryons). In particular,
one finds in both cases a geometric transition from a state in which the 
vacuum forms a finite part of the system to one in which only isolated 
bubbles of vacuum remain in a world of fully or partially overlapping
hadrons. We shall consider this point of ``disappearance'' of a large-scale
vacuum as the end of hadronic matter and calculate the corresponding
limiting curve in the $T-\mu$ plane. 

\medskip

Our question therefore addresses two distinct situations. We have to
consider the percolation of mesons, allowing full overlap, and then
that of baryons with a hard core. In the next section, we shall first 
recall the salient features of percolation theory for the two cases. 
Following this, we will use a very simplistic toy model to illustrate 
the underlying concepts. In section 4, we then turn to the realistic 
case of a hadronic resonance gas with baryon number and strangeness 
conservation. In the last section, we consider the nature of the 
transition and compare our conclusions to the results obtained in 
chiral model studies.

\section{Percolation}

Here we briefly recall the essential results for the percolation of 
spheres, in three space dimensions, for the case of arbitrary overlap
\cite{Isi} and for spheres with an impenetrable hard core, allowing only 
partial overlap \cite{Kratky}. Consider $N$ spheres of radius $R_0$ and 
hence volume $V_0=(4\pi/3)R_0^3$ in a ``box'' of size $V$, with $V \gg V_0$. 
Percolation is said to occur when a connected and hence at least partially 
overlapping set of spheres spans the volume, or in other words, when
the volume occupied by the largest cluster of connected spheres reaches 
a finite fraction of $V$. 

\medskip

We address first the case of permeable spheres, i.e., with arbitrary 
overlap. In this case, when the density $n=N/V$ increases, clusters of
overlapping spheres form, and for
\be
n_m = {\eta_m \over V_0}
\label{p1}
\ee
with $\eta_m \simeq 0.35$, the largest cluster first spans the system,
i.e., percolation occurs \cite{Isi}. At this point, the fraction
\be
\phi_m = 1 - e^{-\eta_m} \simeq 0.30
\label{p2}
\ee
of space is occupied by spheres; the complementary 70\% remain empty 
space (``vacuum''), which also still spans $V$. The largest cluster
therefore has a density of about $1.2/V_0$. In other words, at the 
percolation point, a randomly created state of a system of $N$ spheres 
is very inhomogeneous, consisting mainly of one dense cluster and much 
empty space. A random distribution of extended constituents thus leads 
to something like a ``geometric attraction'' between pointlike constituents, 
with clustering as the result. We shall see later that in the percolation of
hard-core spheres, this attraction is competing with an intrinsic
repulsion. 

\medskip

For $n < n_m$, the clusters of overlapping spheres form only isolated 
bubbles in $V$. From eq.\ (\ref{p1}) we see that at the percolation point,
the total volume of all spheres adds up to 35\% of the total volume; this 
is covered by the spheres to only 30\%, indicating 5\% ``overlap''. After 
a further increase of density, a second percolation point is reached at
\be
n_v =   {\eta_v \over V_0},          
\label{p3}
\ee
where $\eta_v \simeq 1.22$. Now
\be
\phi_v = 1 - e^{-\eta_v} \simeq 0.70
\ee
is the fraction of space covered by spheres, with only 0.30 remaining
empty. Here the sum of the sphere volumes is with 1.22 $V$ considerably
larger than the total volume, due to increased overlap.
For $n \leq n_v$, the vacuum percolates, above $n_v$, only 
isolated bubbles of vacuum remain. In other words, for $ n> n_v$, the 
vacuum has ``disappeared'' as a large-scale entity.

\medskip

The existence of two percolation thresholds, one for the formation of
the first spanning cluster of spheres and another for the disappearence
of a spanning vacuum ``cluster'', is a general feature of percolation
theory in three or more dimensions. 

\medskip

We now turn to the percolation of spheres (again of radius $R_0$) having 
an impenetrable spherical hard core \cite{Kratky}, which we assume to have 
the radius $R_c = R_0/2$. Each sphere now defines a volume 
$V_0=(4\pi /3)R_0^3$ which is not accessible to the center of any other 
sphere. The spheres can partially overlap: the distance between their 
centers has to remain greater than $R_0=2~\!R_c$. With increasing density, 
we now have again two percolation thresholds. At
\be
\n_m = {\bar{\eta}_m \over V_0} 
\label{p4}
\ee  
the spheres form a spanning cluster, and at
\be
\n_v = {\bar{\eta}_v\over V_0}  
\label{p5}
\ee 
the vacuum last spans the volume. Numerical studies \cite{Kratky}
show that $\bar{\eta}_m \simeq 0.34$, practically the same as found above 
for permeable spheres of the same size. The percolation threshold for 
spheres with a hard core is on the dilute side thus not much affected 
by the presence of the hard core, and we have a similar geometric
attraction. The vacuum percolation threshold, 
however, now is given by $\bar{\eta}_v \simeq 2.0$, in contrast to 
$\eta_v \simeq 1.24$ for permeable spheres. In other words, the 
disapperance of the vacuum requires for hard core spheres a higher 
density than needed for permeable spheres. This is directly related 
to the hard core repulsion, which tends to move the spheres apart
and thus at high density counteracts tight clustering.

\section{A Toy Model}

In our simplified toy model, we want to compare the percolation behaviour 
of an ideal gas of massless pions of radius $R_h$ to that of an ideal gas
of massive nucleons of the same size, but having a hard core of a (smaller) 
radius $R_c = R_h/2$. The density of pions is specified by the temperature 
$T$ of the medium, that of the nucleons by the temperature $T$ and the 
baryochemical potential $\mu$, fixing the overall baryon number (nucleons 
minus anti-nucleons).

\medskip

Consider first the system of pions. With three charge states, the 
density is given by
\be
n_{\pi}(T) = 3~ {\pi^2 \over 90}~ T^3.
\label{t1}
\ee
With increasing temperature, they will overlap and eventually fill the 
given volume with one big connected bag. When only isolated vacuum bubbles 
remain, we assume the system to have reached the limit of pionic matter. 
In the previous section it was shown that this occurs for
\be
n_f= {1.22 \over V_h} \simeq 0.57~{\rm fm}^{-3},
\label{t2}
\ee
where $V_h=(4 \pi/3) R_h^3$ and we have used $R_h=0.8$ fm. Solving 
$n_h(T)=n_f$ yields 
\be
T_{\pi} \simeq {0.96 \over R_h} \simeq 240 ~{\rm MeV} 
\label{t3}
\ee 
as the limiting temperature obtained through pion fusion. This value 
will drop slightly when we include nucleons and antinucleons, and
it will decrease considerably when resonances are brought in, as we shall 
see in the next section.

\medskip

The density of pointlike nucleons of mass $M$ is at $T=0$ (when there are no 
anti-nucleons) determined in terms of the baryochemical potential $\mu$, 
\be
n_b(\mu,T=0) = {2\over 3 \pi^2} ~(\mu^2 - M^2)^{3/2}.
\label{t6}
\ee
In the presence of a hard core of volume $V_c=V_h/8$, the density has to be 
calculated taking into account the reduction of the available volume. 
We approximate this by a van der Waals approach \cite{Esko}, 
taking\footnote{A thermodynamically consistent implementation of
hard core repulsion requires in addition a shift of $\mu$ \cite{Rischke},
which for simplicity is neglected here.}
\be
n_B(\mu,T=0) = {n_b(T,\mu) \over 1 + n_b(T,\mu) V_e}
\label{t7}
\ee
for the density of extended nucleons, where $V_e \simeq 2 V_c$ denotes 
the excluded volume at random dense packing of hard spheres \cite{O-L}.
With increasing nucleon density, the box becomes more and more filled, 
and we saw above that the empty vacuum disappears for
\be
{\bar n}_v \simeq {2\over V_h} \simeq 0.93 ~{\rm fm}^{-3}, 
\label{t8}
\ee
taking the same radius for nucleons as for pions; this corresponds to
about 5.5 times standard nuclear density $\rho_0 \simeq 0.17$ fm$^{-3}$.
Solving $n_B(T=0,\mu)={\bar n}_v$ yields 
\be
\mu_v \simeq 1.12 ~ {\rm GeV}
\label{t9}
\ee
for the limiting baryochemical potential at $T=0$. 

\medskip

In the region of low to intermediate $\mu$ in the $T-\mu$ diagram, we 
approximate the density of pointlike nucleons by the Boltzmann limit
\be
n_b(\mu,T)~ \simeq~ {2 T^3 \over \pi^2} \left({M\over T}\right)^2
K_2(M/T) e^{\mu/T}~
\simeq~ {T^3 \over 2} \left({2M \over \pi T}\right)^{3/2}
e^{(\mu - M)/T},
\label{t10}
\ee
where we have also used the large $M/T$ form of the Bessel function
$K_2(M/T)$. At $\mu=0$, we can use this to add nucleons and antinucleons 
to the pions considered so far for filling the box. From
\be
n_{\pi} + 2~\!n_B = {1.22 \over V_h}
\label{11}
\ee
we then find a limiting temperature of hadronic matter
\be
T_h \simeq 230~\!{\rm MeV}
\label{12}
\ee
slightly lower than for pions alone. The small change is due to the 
fact that nucleons amount to only about 6\% of all hadrons at $\mu=0$.
If we would there want to fill the box with only nucleons and antinucleons,
we would have to go to the much higher temperature of about 430 MeV.

\medskip

\begin{figure}[htb]
\centerline{\psfig{file=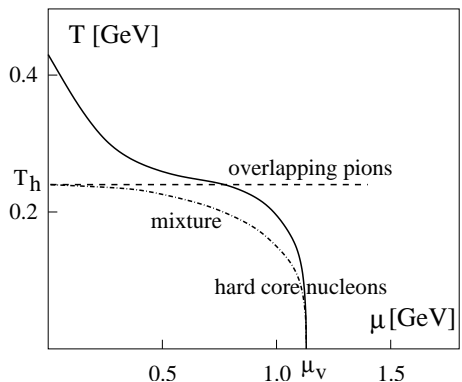,width=6cm}}
\caption{Limit from massless pion percolation and from that of
massive hard core nucleons; the curve labelled ``mixture'' indicates
the result expected of a combined percolation study.}
\label{toy}
\end{figure}

Using the approximation (\ref{t10}), we can ask for the value of $\mu$ at
which the nucleon percolation curve crosses the hadronic limit $T_h$.
Solving $n_B(T=T_h,\mu)=2/V_h$ for $\mu$, we find $\mu\simeq 0.8$ GeV 
for the crossing point. The comparitive behaviour of the two curves is 
illustrated in Fig.\ \ref{toy}. It is evident that at low nucleon density, 
pion percolation limits the hadronic matter density, while at high nucleon 
density, the percolation of hard spheres and the resulting jamming provide 
the limit. It should be emphasized, however, that the separate calculation 
of a (permeable) pion and a (hard-core) nucleon curve is obviously not 
the final solution, since a dilute admixture of nucleons and antinucleons
contributes in the low density regime, just a pions will help to reach
percolation in the nucleonic region. We have followed an additive picture 
here, since so far (to our knowledge), continuum percolation studies have 
been performed either for permeable spheres or for spheres with a hard
core. Evidently what is needed here is a study allowing both types
of constituents in a degree of mixture specified by $\mu$, and this will
lead to modifications in the intermediate region, as schematically
included in Fig.\ \ref{toy}. 

\medskip

\medskip

\begin{figure}[htb]
\centerline{\psfig{file=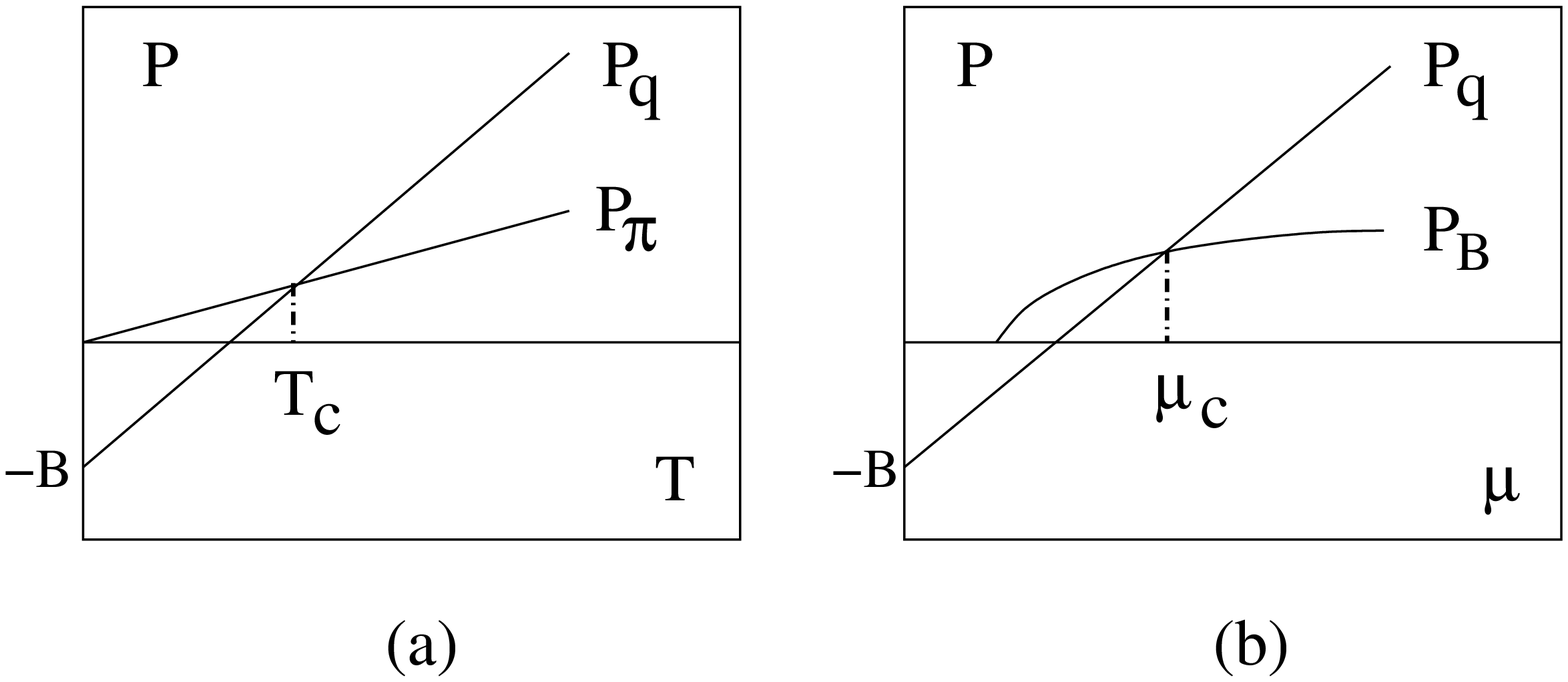,width=8cm}}
\caption{Pressure comparison at $\mu=0$ (a) and at $T=0$ (b)}
\label{bones}
\end{figure}

We note here briefly that we can determine a confinement-deconfinement
limit also by comparing the hadronic pressure to that of 
a quark-gluon plasma \cite{Esko,CGS}.
Such a comparison leads by construction to a first-order transition
in the entire $T-\mu$ plane. For $\mu=0$, we then have
\be
P_{\pi} = 3~ {\pi^2 \over 90}~ T^4.
\label{t13}
\ee
for the pressure of an ideal massless pion gas, to be compared to
\be
P_q = 37~ {\pi^2 \over 90}~ T^4 - B.
\label{t14}
\ee
for that of an ideal massless plasma of two quark flavours; here
$B$ denotes the bag pressure specifying the difference between
coloured and physical vacua. Equating the two pressures 
(see Fig.\ \ref{bones}a) gives
\be
T_c(B)= \left({90 \over 34 \pi^2}\right)^{1/4} B^{1/4}, 
\label{t15}
\ee 
so that by suitably tuning the bag pressure, we can obtain reasonable
values for the transition temperature $T_c$. 

\medskip

For $T=0$, the pressure of extended nucleons is given by
\be
P_B(\mu,T=0) = {P_b(T=0,\mu) \over 1 - n_B(T=0,\mu) V_e},
\label{t16}
\ee
where $n_B$ denotes the corresponding density (eq.\ (\ref{t7})) and $P_b$ 
the pressure of pointlike nucleons of mass $M$ \cite{CGS},
$$
P_b (\mu,T=0) = 
$$
\be
\left({\mu^4 \over 6 \pi^2}\right)
\left\{\left[1 - \left({M\over \mu}\right)^2\right]^{1/2} 
\left[1 - {5 \over 2}\left({M\over \mu}\right)^2\right]
+{3\over 2}\left({M \over \mu}\right)^4 \ln\left[\left({\mu \over M}\right)
\left(1 + \left[ 1- {M^2 \over\mu^2}\right]^{1/2} \right) \right] \right\}
\label{17}
\ee
The corresponding expression for massless quarks is
\be
P_q(\mu,T=0) = {1\over 2 \pi^2} \left({\mu \over 3}\right)^4 - 
B,
\label{18}
\ee
where we have converted the quark baryochemical potential to that
of nucleons, $\mu_q = \mu/3$. Again we equate the two pressures
(see Fig.\ \ref{bones}b) to obtain a critical baryochemical potential
$\mu_c(B,V_c)$. Here as well one can try to tune bag pressure and
hard core volume to obtain a reasonable threshold. 

\medskip

\medskip

However, our point in presenting such toy model hadron-quark pressure
comparison is not a quantitative determination of $T_c(B)$ and 
$\mu_c(B,V_c)$. We only want to illustrate that here in the low 
baryon density region, the confining bag pressure (as counterpart of
the hadron size in percolation) provides the transition and sets 
the transition scale, whereas at low temperature and high baryon density, 
this is achieved by two competing effects, the confining bag pressure 
and the hard-core repulsion.

\medskip

Obviously, these consideration are a gross oversimplification, since
we have neglected the resonance interactions of both pions and nucleons, 
as well as the role of strange mesons and baryons. We shall include these 
in the following. 
 
\section{The Hadronic Resonance Gas}

For vanishing or low baryon number density, the interactions in a hot 
hadronic medium are resonance dominated, and hence the system can be 
described as an ideal gas of all possible resonance species
\cite{B-U,Hage}, contained in an overall spatial volume $V$. 
The grand canonical partition function for such a gas is given by
\cite{KR,KRT} 
\be \ln~Z(T,\mu,\mu_S,V) = 
\ln~Z_{M}(T,V,\mu_S)~+~ \ln~Z_{B}(T,\mu,\mu_S,V)~
 \label{1a} \ee
where the first term gives the meson and the second the baryon contributions 
to the partition function. Baryon number and strangeness are accounted for 
by the corresponding chemical potentials $\mu$ and $\mu_S$, respectively. 
For the meson contributions, we sum over all possible states $i$
\be \ln~Z_{M}(T,V,\mu_S) = \sum_{\rm mesons~i} 
\ln~Z^i_{M}(T,V,\mu_S), \label{1b} 
\ee
with 
\be \ln~Z_{m}^i(T,V,\mu_S) = d_i~ {VT \over2 \pi^2}~ 
m_i^2 \sum_{n=1}^{\infty} n^{-2} K_2(n m_i/T)
~\e^{ nS_i\mu_S/T} , \label{1c} \ee
where $d_i$ specifies the spin-isospin degeneracy  and $S_i$ the 
strangeness of the state $i$.  For the baryon contributions we have
 \be \ln~Z_{B}(T,\mu,\mu_S,V) =   \sum_{\rm baryons~i}
\ln~Z^i_{B}(T,\mu,\mu_S,V), \label{1d} \ee
with
 \be
\ln~Z^i_{B}(T,\mu, V) = d_i~ {VT \over 2
 \pi^2}~ 
m_i^2 \sum_{n=1}^{\infty} (-1)^{n+1}~n^{-2}~K_2(n
m_i/T)~\e^{n(B_i\mu+S_i\mu_S)/T}, \label{1e} \ee

where $B_i$ is the corresponding baryon number of the state $i$. The above 
form incorporates Bose-Einstein and Fermi-Dirac statistics; the first term 
of the sums in Eqs. (\ref{1c}) and (\ref{1e}) is the Boltzmann limit.
In all actual calculations we enforce both baryon number and strangeness
conservation, with a vanishing overall strangeness.

\medskip

The resonances are confined (colour singlet) $\q$ or $qqq$ states of 
hadronic spatial size. As above, we consider the transition from a 
confined to a deconfined medium to occur when the hadrons as little 
bags fuse into one big bag, the quark-gluon plasma \cite{Baacke}; 
the critical transition density is given by eq.\ (\ref{t2}).
If we ignore for the moment the contribution of baryons and antibaryons, 
we can estimate the transition curve in the resonant, low baryon-density 
region by limiting the meson density $n_M$, obtained from eq.\ (\ref{1c}),
\be
n_M(T,\mu) = {1.22 \over V_h} \simeq 0.57~{\rm fm}^{-3}.
\label{3}
\ee
This equation can be solved using a resonance gas summation code,
summing over all meson states up to mass 2.5 GeV \cite{KR}; it leads 
for $\mu=0$ to a transition temperature $T_c \simeq 177$ MeV, 
a value considerably lower than that of pure pion gas case in the
previous section

\medskip

Just as in the toy model, however,
the temperature obtained from eq.\ (\ref{3}) is presumably somewhat 
too high, since we had required the box to be filled by mesons 
alone, while in fact baryons and antibaryons can contribute to forming a 
connecting cluster of hadronic matter. But including the density 
of baryons and antibaryons in the limiting relation (\ref{3}) gives 
at $\mu=0$ a limiting temperature $T_c \simeq 170$ MeV, only slightly 
lower, so that the role of the baryons and antibaryons in establishing
the boundary is again quite small.

\medskip

\begin{figure}[htb]
\centerline{\psfig{file=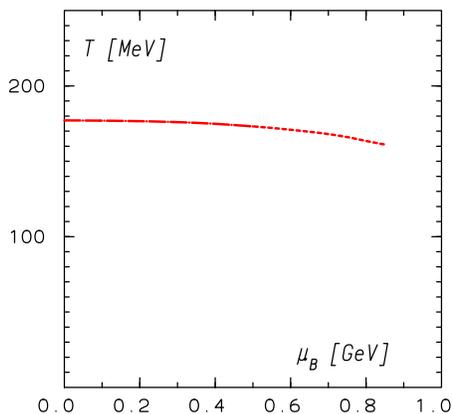,width=6cm}}
\vskip-1cm
\caption{Transition line in the resonance regime}
\label{bag}
\end{figure}

If we neglect the interrelation of strangeness and baryon number due
to associated production, the mesonic part of the partition function
is independent of $\mu$, and hence eq.\ (\ref{3}) leads to a constant 
$T_c$ for all $\mu$. Taking associated production into account, however, 
implies with increasing $\mu$ an increasing density of strange mesons and 
thus a (slightly) decreasing temperature. The resulting behaviour is
shown in Fig.\ \ref{bag}. 

\medskip

\vskip-1cm

The deconfinement transition in the resonant region is  
conceptually a consequence of hadron size and clustering, leading to
the fusion into one connected volume \cite{Baacke}, and it occurs, 
as we saw above, already for a pion gas, though at a higher temperature.
Thus the existence of the limit is due to the basic hadronic size 
\cite{Pomer}, the actual value of the transition temperature to the 
scale specifying the resonance spectrum \cite{Hage}. 

\medskip

An interacting hadron medium can be replaced by an ideal resonance gas
only if the interactions are dominantly of resonance nature. As the baryon
density increases, however, there are more and more non-resonant contributions,
and in the limit of low temperature $T$ and large baryochemical potential 
$\mu$, i.e., for cold and dense nuclear matter, the non-resonant nuclear
forces are the dominant interaction.

\medskip

\begin{figure}[htb]
\vskip-1cm
\centerline{\psfig{file=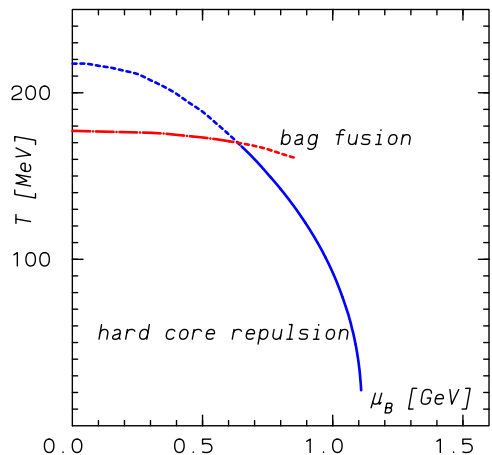,width=6.5cm}}
\vskip-1.3cm
\caption{Bag fusion vs. hard core transition lines}
\label{bag-core}
\end{figure}

We therefore assume as above that in the region of large baryon
density, the jamming of baryons with a hard core defines the limit,
so that the relevant relation is eq.\ (\ref{t8}). To connect this 
geometric argument with thermodynamics, we again follow the van der 
Waals approach (\ref{t7}) for baryons and antibaryons, 
with $V_e=2V_c$ and $R_c=0.4$ fm. The 
resulting hard core transition curve is 
shown in Fig.\ \ref{bag-core}, together with the bag fusion curve. 
The $T-\mu$ plane thus shows two distinct regimes: at low $\mu$, 
hadron percolation through bag fusion, at large $\mu$ a first-order 
mobility or jamming transition. 

\medskip

As already mentioned above, the separate calculation of a bag fusion 
and a hard core curve is only an attempt to arrive at a schematic
picture as long as we don't have continuum percolation studies for
a ($\mu$-determined) mixture of permeable spheres with spheres having  
a hard core. The curve expected from such a calculation will presumably
lead to a transition from one regime to the other at somewhat lower
values of $\mu$. In fact, the baryon density becomes equal 
to the meson density at $\mu \simeq 0.45$ GeV, and therefore this may 
well be close to the point at which the change in the nature of the 
transition occurs.

\section{The Nature of the Transition}

Our considerations up to now specified the limit of hadronic matter,
defined as the point of disappearance of the vacuum as a large-scale 
feature. This point was determined through percolation studies, and
the percolation limit is in general not a thermodynamic phase transition.
Percolation can thus naturally provide a way to produce a rapid cross-over not 
associated with any singularity of the partition function \cite{S-perc,F-S}.
It should be noted, however, that for spin systems, thermal critical behaviour 
can be formulated in terms of percolation \cite{F-K,C-K,Blan}.
It seems possible to extend this to gauge systems, and first such studies
relate the onset of deconfinement at $\mu=0$ to Polyakov loop 
percolation \cite{Polyaperc}, analogous to the onset of magnetisation 
as the percolation of spin clusters. Here it is the onset of large 
scale disorder, i.e., of the vacuum, which induces critical behaviour.
From the confined side, we thus have hadronic bag fusion leading to
the disappearence of the physical vacuum, while on the deconfined side,
formation of disordered clusters in an ordered medium correspond to the
appearence of the vacuum. Based on the spin-gauge universality \cite{S-Y}, 
deconfinement as Polyakov loop percolation could occur as first (SU(3)) 
or second order (SU(2)) phase transition, corresponding to the spontaneous 
breaking of a global center $Z_3$ or $Z_2$ symmetry.  

\medskip

In the case of hard core percolation, a connection to thermodynamic 
critical behaviour has also been discussed \cite{Kratky}. If a system 
with hard core repulsion between its constituents is in addition
subjected to a density-dependent negative background potential, first
order critical behaviour can appear. A classical case is the van der
Waals equation. The pressure in our hard core medium is given by
\be
P(T,\mu) = {P_0(T,\mu) \over 1 - n(T,\mu) V_e}, 
\label{n1}
\ee
where $P_0$ denotes the pressure of pointlike constituents. The
density $n$ of the hard core constituents is given by eq.\ (\ref{t7}),
and $V_e$ again denotes the random dense packing volume. If we add
to this purely repulsive form a density dependent attractive term 
\be
P_{\w}(T,\mu) = P(T,\mu) - a~\!n^2,
\label{n2}
\ee
with constant $a$,
we obtain the van der Waals equation of state with a first order phase 
transition ending in a second order critical point specified by the 
parameters $a$ and $V_e$.

\medskip

Let us now recall how the transition is treated in the Nambu-Jona-Lasinio
model (NJL). Here the thermodynamic potential $\Omega(m)$ depends on the 
dynamically generated mass $m \simeq \langle \bar \psi \psi \rangle$, 
which serves as an order parameter for chiral symmetry. By analyzing the 
minimum conditions, 
\be
{\partial~\! \Omega \over \partial m} = 0 ~~~~
{\partial^2 \Omega \over \partial^2 m} > 0,
\ee
at fixed $T$ and $\mu$, one studies the onset of chiral symmetry breaking 
and the order of the transition. At $T=0$ and $\mu=0$, the mass is generated 
through the attractive self-energy interaction 
$\sim G (\bar \psi \psi)^2$, which results in chiral 
symmetry breaking. In a thermal medium of vanishing baryon density, an 
increase in the temperature produces fluctuations which eventually overcome 
the attraction and thus lead to the chiral symmetry restoration transition.
At large baryon density and low temperature, the attraction has to compete
with the repulsion due to the Fermi statistics of the quarks. As
above, this competition brings about a first order phase transition,
which persists in some temperature range $0 < T \leq T_c$; here
$T_c$ is essentially the point at which the Fermi repulsion is
fully overcome by the thermal fluctuations. The mechanism for
the large $\mu$ behaviour in the NJL model is
thus conceptually very similar to a van der Waals pattern \cite{Iwasaki}.
Note that in addition to the mentioned mechanisms, at sufficiently low
$T$ there will presumably be $qq$ pairing into bosonic diquarks,
leading to condensation and colour superconductivity.
We do not consider this here, although it can certainly modify the
resulting phase structure \cite{Baym}. 

\medskip

Since the NJL model does not contain gluons, it also does not include
any confinement-deconfinement transition. The latter is implemented
by introducing an interaction term with Polyakov loops \cite{WW,Gatto}. 
The resulting description leads in the $T\!-\!\mu$ plane to a deconfinement 
curve determined by the Polyakov loop contribution, which at some (large)
$\mu$ intercepts the chiral curve \cite{CS,Gatto}, in a form very similar to 
that of Fig.\ \ref{bag-core}. 

\section{Summary}

We have argued that as function of $T$ and $\mu$, hadronic matter
finds itself in two distinct regimes. At low baryon density, the
behaviour of the system is governed by resonance formation and 
clustering, with hadronic size and resonance spectrum as relevant
parameters. The interaction here is essentially attractive. At
high baryon density, in addition to this, there is a repulsive
contribution, on the confined side as nuclear repulsion, on the
deconfined side as Fermi repulsion between quarks. In our approach, 
the resulting limit of hadronic matter is in the low baryon density 
region determined by the percolation of permeable (overlapping) mesons 
and at high baryon density by the percolation of hard-core baryons. 
In the latter case, the competition between repulsion and clustering 
can provide  a first order phase transition. In mesonic percolation 
there is only clustering; while in general not resulting in thermodynamics 
critical behaviour, it can in specific cases (depending on details of
the dynamics) also result in first or second order phase transitions.  

\bigskip

\centerline{\bf \large Acknowledgments}

\medskip

We thank F.\ Karsch for helpful remarks. K.R. also acknowledges 
fruitful discussions with P. Braun-Munzinger, and thanks the Polish 
Ministry of Science and the Deutsche Forschungsgemeinschaft (DFG) 
under the Mercator Programme for partial support.


\bigskip


\begin{thebibliography}{99}

\bibitem{Karsch} For a recent survey, see e.g., F.\ Karsch,
J.\ Phys.\ Conf.\ Ser.\ 46 (2006) 122.

\bibitem{Fodor} Z.\ Fodor and S.\ Katz, JHEP 0203 (2002) 014.

\bibitem{Forcrand} P.\ de Forcrand and O. Philipsen, \NP B 642 (2002) 290.

\bibitem{Lombardo} M.-P.\ Lombardo, \PR D 67 (2003) 014505.

\bibitem{Swansea} C.\ R.\ Allton et al., \PR  D 68 (2003) 014507;\\
C.\ Miau and C.\ Schmidt, PoS (LATTICE 2007) 175.

\bibitem{Stepha} M.\ A.\ Stephanov, \PRL 76 (1996) 4472.

\bibitem{Halasz} M.\ Halasz et al., \PR D 58 (1998) 096007.

\bibitem{SRS} M.\ A.\ Stephanov, K.\ Rajagopal and E.\ Shuryak, \PRL 81
(1998) 4816

\bibitem{SKP} T.\ M.\ Schwarz, S.\ P.\ Klevansky and G.\ Papp, \PR C 60 (1999)
055205.

\bibitem{Buballa} M.\ Buballa, Phys.\ Rep.\ 407 (2003) 205. 

\bibitem{CS}
C.~Sasaki, B.~Friman and K.~Redlich,  Phys.\ Rev.\  D 75 (2007) 074013.

\bibitem{WW}
P.\ N.\ Meisinger and M.\ C.\ Ogilvie, \PL B 379 (1996) 163;\\
K.\ Fukushima, \PL B 591 (2004) 277;\\
C.~Ratti, M.~A.~Thaler and W.~Weise, Phys.\ Rev.\ D 73 (2006) 014019.

\bibitem{KS} See e.g., F.\ Karsch and H.\ Satz, \PR D 21 (1980) 1168.

\bibitem{Isi} See e.g., M.\ B.\ Isichenko, Rev.\ Mod.\ Phys.\ 64 (1992) 961.

\bibitem{Kratky} K.\ W.\ Kratky, J.\ Stat.\ Phys.\ 52 (1988) 1413.

\bibitem{Esko} J.\ Cleymans et al., \ZP C 33 (1986) 151.

\bibitem{Rischke} D.\ H.\ Rischke et al., \ZP C 51 (1991) 488;\\
G.\ D.\ Yen et al., \PR C 56 (1997) 2210.

\bibitem{O-L}  G.\ Y.\ Onoda and E.\ G.\ Liniger, \PRL 22 (1990) 2727.

\bibitem{CGS} J.\ Cleymans, R.\ V.\ Gavai and E.\ Suhonen, 
Phys.\ Rept. 130 (1986) 217. 

\bibitem{B-U} E.\ Beth and G.\ E.\ Uhlenbeck, Physica 4 (1937) 915.

\bibitem{Hage} R.\ Hagedorn, Nuovo Cim. Suppl. 3 (1965) 147;\\
Nuovo Cim. 56 A (1968) 1027.

\bibitem{KR}  P. Braun-Munzinger, K. Redlich and J. Stachel,  
``Quark Gluon Plasma 3",
 Eds. R.C. Hwa and Xin-Nian Wang, World Scientific Publishing (2003) 491.

\bibitem{KRT} F.\ Karsch, K.\ Redlich and A.\ Tawfik, \PL B 571 (2003) 67.

\bibitem{Baacke}  J.\ Baacke, Acta Phys.\ Polon.\ B 8 (1977) 625.

\bibitem{Pomer}  I.~Ya.~Pomeranchuk, Doklady Akad. Nauk. SSSR 78 (1951) 889.

\bibitem{S-perc} H.\ Satz, \NP A 642 (1998) 130c.

\bibitem{F-S} S.\ Fortunato and H.\ Satz, \PL B 509 (2001) 189.

\bibitem{F-K} C.\ M.\ Fortuin and P.\ W.\ Kasteleyn,
J.\ Phys.\ Soc.\ Japan {26} (Suppl.) (1969) 11;\\
Physica {57} (1972) 536.

\bibitem{C-K} A.\ Coniglio and W.\ Klein, J.\ Phys.\ A {13} (1980) 2775.

\bibitem{Blan} Ph.\ Blanchard et al., J.\ Phys.\ A 41 (2008) 085001.

\bibitem{Polyaperc}  S.\ Fortunato and H.\ Satz, \PL B 475 (2000) 311.

\bibitem{S-Y} B.\ Svetitsky and L.\ G.\ Yaffe, \NP B 210 [FS6]
(1982) 423.

\bibitem{Iwasaki} M.\ Iwasaki, arXiv:hep-ph/0411199, Nov. 2004

\bibitem{Baym} See e.g., G.\ Baym et al., \PR D 76 (2007) 074001;\\
L. McLerran and R.\ D.\ Pisarski, \NP A 796 (2007) 83.
 
\bibitem{Gatto} H.\ Abuki et al., arXiv:0805.1509v2 [hep-ph], May 2008.



\end{thebibliography}
\end{document}